\documentstyle[11pt]{article}
\title{\begin{flushright}
{\normalsize
NSF-ITP-94-28\\
NUC-MINN-94/3-T \\
April 1994\\}
\end{flushright}
\bf THE PROXIMAL CHIRAL\\ PHASE TRANSITION}
\author{ {\bf Joseph I. Kapusta} \\
  {\small \it School of Physics and Astronomy, University of Minnesota,
Minneapolis, MN 55455}\\
{\bf Ajit M. Srivastava} \\
{\small \it Institute for Theoretical Physics, University of California,
Santa Barbara, CA 93106}}

\date{}

\begin{document}

\maketitle

\begin{center}
Abstract\\
\end{center}

We consider the form of the chiral symmetry breaking piece of the effective
potential in the linear sigma model.  Surprisingly, it allows for a second
local minimum at both zero and finite temperature.  Even though chiral
symmetry is not exact, and therefore is not restored in a true phase
transition at finite temperature, this second minimum can nevertheless
mimic many of the effects of a first order phase transition.  We derive a
lower limit on the height of the second minimum relative to the global
minimum based on cosmological considerations; this limit is so weak as
to be practically nonexistent.  In high energy nuclear collisions, it
may lead to observable effects in Bose-Einstein interferometry due to
domain walls and to coherent pion emission.

\vfill \eject

\section{Introduction}

The possibility of producing quark-gluon plasma in
relativistic heavy ion collisions is an
exciting one, especially from the point of view of observing the
chiral/confinement phase transition/crossover as the
plasma expands and cools. Formation of domains in a chiral phase
transition where the chiral field may not be oriented along the
true vacuum has been a subject of many investigations recently.
Formation of a large domain with a {\em Disoriented Chiral Condensate}
(DCC) has been proposed by Bjorken, Kowalski and Taylor
\cite{bkt} in the context of high multiplicity hadronic collisions.
It was argued in \cite{bkt} that, as the chiral field relaxes
to the true vacuum in such a domain, it may lead to coherent emission
of pions \cite{bd}. A motivation for this proposal comes from
Centauro  events in cosmic ray collisions \cite{cntr}.
In the context of quark-gluon plasma, Rajagopal and Wilczek
proposed \cite{rw} that the nonequilibrium dynamics during the phase
transition may produce  DCC domains. They argued that long wavelength
pion modes may get amplified leading to emission of coherent pions.

One difficulty in these scenarios is that one typically expects
domains which are not much bigger than the pion size \cite{ggp}.
Several studies have focussed on the possibility of getting a larger
domain.  Gavin, Gocksch and Pisarski have argued
\cite{ggp} that large domains of DCC can arise if the effective masses
of mesons are small, while Gavin and M\"uller propose \cite{lrg} the annealing
of smaller domains to give a large region of DCC.

In this paper we consider the structure of the effective potential
in the chiral model and note that more general possibilities exist
for the symmetry breaking term than considered in these previous
investigations.  {\em This allows
for the existence of a second local minimum of the potential, in addition
to the true global minimum, leading to the formation of
domain walls which interpolate between the two minima.}
These walls are unstable unless the two minima
are exactly degenerate. We consider constraints coming from cosmology
on the parameters responsible for the existence of such walls. We find
that such constraints are extremely weak. We then consider
phenomenological consequences of the richer structure of the effective
potential of the model, especially from the point of view of the
formation of DCC.  We show that in these more general models large
domain walls naturally form but
eventually disappear, leading to emission of pions from shell-like
structures. Bose-Einstein interferometry should be able to reveal any
such shells \cite{pratt}. We further argue that large regions of
DCC may arise naturally in these models, and may be able to account
for phenomena like Centauro events.

There is some analogy to the physics of spin glasses \cite{glass}.
A spin glass is characterized by a phase space which has a complicated
landscape of valleys.  As the temperature is reduced, barriers between
valleys become significant, and the relaxation times become very long.
The system may even get trapped in a single metastable state for
the duration of the experiment.  One may observe hysteresis in strong
magnetic fields that varies with the experimental conditions.
Real spin glasses have an anisotropy.  In hydrodynamic models of
Heisenberg spin glasses one is naturally led to consider a coarse-grained
free energy which has similarities to the effective potential given
in eq. (20) below.

The paper is organized as follows. In sect. 2 we consider the
details of the effective potential, both at zero temperature
and at finite temperature, and discuss various
possibilities for the symmetry breaking term. In sect. 3 we
consider constraints on the parameters of the model arising from
the formation of domain walls in the early Universe.  In sect. 4
we consider potentially observable effects in high energy collisions,
namely, pion production from domain walls.  In sect. 5 we present
some conclusions.

\section{The Effective Potential}

The sigma model \cite{gml} is thought to represent the
long wavelength limit of QCD \cite{weinb,gasl}.  Much has been
written about the model, both at zero temperature \cite{lee,donob}
and at finite temperature \cite{wil}.  Despite this, the shape
of the effective potential at both zero and finite temperature
may have a nontrivial structure when the pion mass is nonzero,
which feature seems to have escaped attention; at least
it goes against the conventional considerations.  In this section
we explore the effective potential, and in the following sections
we shall explore its phenomenological consequences.

\subsection{The vacuum}

We use the sigma model in its linear representation.
The Lagrangian is expressed in terms of a scalar field $\sigma$ and
the pion field \mbox{\boldmath $\pi$}.
\begin{equation}
{\cal L} \,=\, \frac{1}{2}\left(\partial_{\mu}\sigma\right)^2
+ \frac{1}{2}\left(\partial_{\mu} \mbox{\boldmath $\pi$}\right)^2
-\frac{\lambda}{4} \left(\sigma^2 + \mbox{\boldmath $\pi$}^2 -
c^2/\lambda\right)^2 - V_{SB} \, .
\end{equation}
The piece of the Lagrangian which explicitly breaks chiral symmetry
is $V_{SB}$.  In the absence of this term, the potential has
the shape of the bottom of a wine bottle.  Chiral symmetry is
spontaneously broken in the vacuum, the pion is the massless Goldstone
boson, and the $\sigma$ meson gets a mass
on the order of 1-2 GeV.  The axial vector current, defined by
\begin{equation}
{\bf A}_{\mu} \,=\, \left(\partial_{\mu}\sigma\right)
\mbox{\boldmath $\pi$}
- \left(\partial_{\mu} \mbox{\boldmath $\pi$} \right)\sigma \, ,
\end{equation}
is conserved.  It is believed that chiral symmetry is restored by a
phase transition, which is perhaps of second order, at a critical
temperature $T_c \approx 160$ MeV; we come back to this point later.

Any quantitative description of Nature at finite momentum and energy
must include the vector mesons \cite{vector}.  Since we are proposing
here a qualitatively new phenomenon we shall neglect them, as well
as strangeness.

The up and down quarks, while very light, are not massless; therefore,
neither is the pion.  Historically, there have been three ways to add
a symmetry breaking term to the linear sigma model \cite{camp}:

\noindent {\bf 1:} $V_{SB} = -f_{\pi}m_{\pi}^2 \sigma$, which is linear in
the $\sigma$ field.  The PCAC relation
\begin{equation}
\partial_{\mu}{\bf A}^{\mu} \,=\, f_{\pi}m_{\pi}^2
\mbox{\boldmath $\pi$}
\end{equation}
is realized as an operator equation.

\noindent {\bf 2:} $V_{SB} = \frac{1}{2} m_\pi^2 \mbox{\boldmath $\pi$}^2$,
which is quadratic in the pion field.  The PCAC relation is realized only
in the weak field limit.

\noindent {\bf 3:} $V_{SB} = {\rm constant}\times{\bar N}N$, which is
quadratic in the nucleon field.  The divergence of the axial vector
current is proportional to the axial vector nucleon current.

In principle one could imagine that low energy pion
scattering measurements could distinguish between (1) and (2).
However, by its very nature that occurs in the weak field limit, so
the measurements would have to be very precise.  Also, low energy
pion scattering is influenced by the tails of resonances, such as the
$\rho$ meson \cite{dono}, which would tend to mask the effects of the
nonlinear symmetry breaking terms.  We are not aware of
any experimental evidence which prefers one symmetry breaking term over
the other.  In this paper we shall not consider the possibility {\bf 3}.

If one insists on an effective Lagrangian which is rotationally
invariant and renormalizable, then the most general symmetry breaking
potential can be written as
\begin{equation}
V_{SB} \,=\, - \sum_{n=1}^4 \frac{\epsilon_n}{n!}\sigma^n
\,+\, (\delta_1\sigma + \delta_2 \sigma^2)
\mbox{\boldmath $\pi$}^2 \,.
\end{equation}
Other symmetry breaking terms one might think of adding simply amount
to a redefinition of the 8 parameters $\lambda, c, \epsilon_n, \delta_n$.
Relaxation of the renormalizability
condition would allow further terms but, as we shall see, there is
already sufficient freedom to generate interesting physics.
In what follows we will set $\delta_1 = \delta_2 = 0$,
mainly for simplicity of presentation.

With the symmetry breaking potential as given above, the divergence
of the axial vector current is
\begin{equation}
\partial_{\mu}{\bf A}^{\mu} \,=\, -V_{SB}'(\sigma)
\mbox{\boldmath $\pi$} \, ,
\end{equation}
where the prime denotes differentiation with respect to $\sigma$.
It is clear that PCAC is an operator identity only if $V_{SB}$ is
linear in $\sigma$.  The ground state of this theory ought to occur
at $\sigma(x) = \sigma_{\rm gs} > 0$ and $\mbox{\boldmath $\pi$}(x) = \bf 0$.
We must immediately impose three conditions; that the minimum of the full
potential occur at $\sigma_{\rm gs}$, that PCAC hold for small fluctuations
about $\sigma_{\rm gs}$, and that the pion has its physical mass there.
Therefore,
\begin{equation}
V_{SB}'(\sigma_{\rm gs}) \,=\, -f_{\pi}m_{\pi}^2 \, ,
\end{equation}
\begin{equation}
\sigma_{\rm gs}(\lambda \sigma_{\rm gs}^2 - c^2) + V_{SB}'
(\sigma_{\rm gs}) \,=\, 0 \, ,
\end{equation}
\begin{equation}
\lambda \sigma_{\rm gs}^2 - c^2 \,=\, m_{\pi}^2 \, ,
\end{equation}
and so $\sigma_{\rm gs} = f_{\pi}$.

One must still ensure that the global minimum is really obtained when
$\mbox{\boldmath $\pi$}(x) = \bf 0$.  To investigate this problem,
let us expand
the fields about an arbitrary point as follows.
\begin{eqnarray}
\sigma(x) &=& v\,\cos\theta + \sigma'(x) \, ,\\
\mbox{\boldmath $\pi$}(x) &=& {\bf v}\,\sin\theta +
\mbox{\boldmath $\pi$}'(x) \, .
\end{eqnarray}
The primes denote fluctuations about the given point.  The full potential
is
\begin{equation}
V(v,\theta) \,=\, \frac{\lambda}{4} \left(v^2 -c^2/\lambda\right)^2
+ V_{SB}(v\,\cos\theta) \, .
\end{equation}
We now consider several limits for the symmetry breaking piece of
the potential.

Suppose that $\epsilon_1 = f_{\pi}m_{\pi}^2$ and all other $\epsilon$'s
are zero.  Minimizing the potential with respect to $v$ at fixed
$\theta$ gives
\begin{equation}
v(\theta) \, \approx \, f_{\pi}(1 + \Delta) \, ,
\end{equation}
where $\Delta = - 2(m_{\pi}/m_{\sigma})^2 \sin^2(\theta/2)$,
$m_{\sigma}^2 = 2\lambda f_{\pi}^2 + m_{\pi}^2$, and
\begin{equation}
V_{\rm min}(\theta) - V_{\rm min}(0) \, \approx \,
2 f_{\pi}^2 m_{\pi}^2 \sin^2(\theta/2) \, .
\end{equation}
Corrections to the approximate equalities are of relative order
$\Delta << 1$.  This is a tilted wine bottle bottom with only one
minimum.

Now suppose that $\epsilon_2 = m_{\pi}^2$ and all other $\epsilon$'s
are zero.  Then
\begin{equation}
v^2(\theta) \,=\, f_{\pi}^2 [1 - 2 m_{\pi}^2
 \sin^2\theta/m_{\sigma}^2] \, ,
\end{equation}
where $m_{\sigma}^2 = 2\lambda f_{\pi}^2$, and
\begin{equation}
V_{\rm min}(\theta) - V_{\rm min}(0) \, \approx \,
\frac{1}{2} f_{\pi}^2 m_{\pi}^2 \sin^2\theta \, .
\end{equation}
We neglect a correction of relative order $(m_{\pi}/m_{\sigma})^2$.
This potential has two degenerate minima located at $\theta = 0$
and $\pi$!

In general, one may expect that $V_{SB}$ allows for two minima, one
at $\theta = 0$ and one at $\theta = \pi$.  If they are not degenerate
then by a simple redefinition of the fields we may, by convention,
choose $\theta = 0$ to be the global minimum.  We shall investigate
what limits cosmology may place on the existence and depth of the
second minimum in the next section; we shall find that the constraint
is extremely weak.  It is quite surprising to us that {\em neither
terrestrial experiments nor pure theoretical computations in QCD so
far tell us anything about a possible second minimum.}

Our statement certainly
goes against the conventional point of view as expounded in ref.
\cite{Pagels}, for example, which says that the symmetry breaking
potential should be linear in the fields, and for {\em three}
quark flavors should follow the $(3,3^*)+(3^*,3)$ symmetry breaking
scheme of Gell-Mann, Oakes and Renner \cite{GOM}.  The main argument
seems to be that this is the simplest description which gives reasonable
low energy phenomenology.  There was some interest in this issue in
the 1970's in regard to three flavor physics.  It was found that the
addition of symmetry breaking terms bilinear in the scalar fields
resulted in low energy phenomenology as good as, or better than,
linear terms alone \cite{bi}.  These bilinear terms could have the
structure $(3,3^*)+(3^*,3)$ or they could have components of some
other group structure.  A more recent study \cite{Schecht} has
found a nonzero coefficient of a bilinear term in the three flavor
{\em nonlinear} sigma model.  We will not pursue the three flavor world
in this paper, but it is certainly worth doing.

One might at first think that $\epsilon_n \propto m_q^n$ (where $m_q$
is the up or down quark mass) so that $\epsilon_1$ is much greater
than $\epsilon_2$ and so on.  We think it is quite possible that all
$\epsilon_n \propto m_q$ and therefore of comparable magnitude
(when scaled appropriately with $f_{\pi}$).  The argument is that
the sigma model is only a low energy effective model of QCD and
all possible terms which are allowed should be included.  Indeed,
even if one started originally with only a linear symmetry breaking
term $\epsilon_1'$, loop corrections would generate nonlinear terms.
These nonlinear terms would have coefficients equal to
$\epsilon_1'$ times some function of $\lambda$ and $c$.  Since
$\lambda$ and $c$ are big we expect that all $\epsilon_n \,
f_{\pi}^n$ would turn out to be comparable in magnitude.

For clarity of exposition we shall hereafter restrict our
attention to the possibility
that only $\epsilon_1$ and $\epsilon_2$ are nonzero.  This is sufficient
to parametrize the effective potential with the freedom to adjust the
tilt of the bottom of the wine bottle as well as the depth of the
second minimum.  We have then at our disposal four parameters in the
effective Lagrangian: $\lambda, c, \epsilon_1$ and $\epsilon_2$.
These parameters must be restricted so as to give the proper pion
mass, pion decay constant, a reasonable value for the $\sigma$ mass,
PCAC in the weak field limit, and the condition that the ground
state of the theory occur at $\sigma = f_{\pi}$ and
$\mbox{\boldmath $\pi$} = {\bf 0}$.  We obtain
\begin{eqnarray}
m_{\pi}^2 &=& \lambda f_{\pi}^2 - c^2 \, ,\\
m_{\sigma}^2 &=& 2 \lambda f_{\pi}^2 + m_{\pi}^2 - \epsilon_2 \, ,\\
f_{\pi} m_{\pi}^2 &=& \epsilon_1 + \epsilon_2 f_{\pi} \, .
\end{eqnarray}
The numerical values chosen in this paper are $m_{\pi} = 140$ MeV
and $f_{\pi} = 94.5$ MeV.  Lin and Serot \cite{lin} have emphasized
that the $\sigma$ meson in this model is {\em not} to be identified
with the exchange of two correlated pions in the isoscalar - scalar
channel in the nucleon - nucleon interaction.  That exchange is
rather broadly distributed in mass with a peak around 600 MeV.
Good phenomenology for low energy pion and nucleon dynamics is
obtained if the $\sigma$ meson has a mass greater than about
1 GeV.  For definiteness, we choose $m_{\sigma} = 1$ GeV.
We shall vary $\epsilon_1$ between 0 and $f_{\pi} m_{\pi}^2$.  There
is no further freedom given the above constraints.

\subsection{Finite temperature}

To estimate what may happen at finite temperature we will calculate
thermal fluctuations to one loop order and furthermore take the
high temperature limit.  Quantitatively this cannot be very accurate.
The relevant coupling constant is large: $\lambda \approx (m_{\sigma}
/f_{\pi})^2/2 \approx 50$.  Qualitatively the result should
be alright; however, see \cite{rw,rob}.

To proceed, we expand the fields about an arbitrary point, as in
eqs. (9-10), and determine the masses of the fluctuations.  If $\bf v$
points in the third direction in isospin space then the eigenvalues
of the mass-squared matrix are
\begin{eqnarray}
m_1^2 &=& \lambda v^2 - c^2 \, , \nonumber \\
m_2^2 &=& \lambda v^2 - c^2 \, , \nonumber \\
m_3^2 &=& 2 \lambda v^2 - c^2 - \frac{\epsilon_2}{2}
- \sqrt{\lambda^2v^4 - \epsilon_2 \lambda v^2 \cos(2\theta)
+\frac{\epsilon_2^2}{4}} \, , \nonumber \\
m_0^2 &=& 2 \lambda v^2 - c^2 - \frac{\epsilon_2}{2}
+ \sqrt{\lambda^2v^4 - \epsilon_2 \lambda v^2 \cos(2\theta)
+\frac{\epsilon_2^2}{4}} \, .
\end{eqnarray}
In the high temperature limit of the one loop approximation one keeps
only the terms of order $T^4$ and $m^2T^2$.  Ignoring terms which
are independent of $v$ and $\theta$ we get the simple expression
\begin{equation}
V(v, \theta; T) \,=\, \frac{\lambda}{4}v^4 - \frac{1}{2}
\left( c^2 + \epsilon_2 \cos^2\theta - \frac{\lambda T^2}{2}
\right) v^2 - \epsilon_1 v \cos\theta \, .
\end{equation}

In the chiral limit one finds, as is well-known, a second order
phase transition at the critical temperature
$T_c = \sqrt{2c^2/\lambda} = \sqrt{2} f_{\pi}$.
An analysis by Karsch \cite{karsch} of all available lattice
simulations of two-flavor
QCD extrapolated to zero quark mass is consistent with a second order
transition with critical indices the same as the O(4) model.  So at
least qualitatively the model and approximations made here make sense.
However, as emphasized by Shuryak \cite{ed}, the sigma model is supposed to
represent only the long wavelength modes, and certainly does not
include the contribution from short wavelength modes.  For example,
as one approaches $T_c$ from below, the model does not include
the $\eta, \rho, \omega$, and the whole tower of mesons above them.
As one approaches $T_c$ from above, the model does not include all the
degrees of freedom represented by quarks and gluons.  The energy density
of the long wavelength modes represented by the pion and sigma degrees
of freedom should be thought of as sitting on top of a much larger
energy density represented by all these other degrees of freedom.

When the up and down quark masses are nonzero chiral symmetry is
not exact.  It cannot be restored at high temperature.  If $V_{SB}$
is an even function of $\sigma$ the Lagrangian still possesses a
discrete symmetry which is restored at some critical temperature.
For example, when only $\epsilon_2 \neq 0$, this temperature is
$\sqrt{2(c^2 + m_{\pi}^2)/\lambda}$.

It is straightforward to show that the zero temperature effective
potential has a second, local, minimum at $\theta = \pi$ which is
separated from the global minimum at $\theta = 0$ by a barrier when
the inequality $\epsilon_2 \sqrt{c^2/\lambda} > \epsilon_1 > 0$ is
satisfied.  [$\sqrt{c^2/\lambda}$ is just $f_{\pi}$ up to corrections
of order $m_\pi^2/m_\sigma^2$.]  As the temperature is increased,
this minimum develops into a saddle at temperature $T_1$ where the
curvature in the azimuthal direction changes from positive to
negative.  The saddle eventually disappears at the higher temperature
$T_2$.  Here
\begin{equation}
T_1 \,=\, \sqrt{2\left[ \frac{c^2}{\lambda}-\left(\frac{\epsilon_1}
{\epsilon_2}\right)^2 \right]} \, ,
\end{equation}
and
\begin{equation}
T_2 \, = \, \sqrt{\frac{2(c^2+\epsilon_2)}{\lambda}
-6\left( \frac{\epsilon_1}{2\lambda} \right) ^{2/3} } \, .
\end{equation}

In figures 1 through 4 we show the evolution of the effective potential
with increasing temperature for four sets of $\epsilon$.
All of these represent a slice through the $V - \sigma$ plane,
and are normalized such that $V_{min} = 0$ at each temperature.
One can imagine that a system cooling through $T_1$ could
get trapped in the metastable minimum.  We shall consider
such possibilities in the following sections.  We call this the
{\em proximal} chiral phase transition, since it is a consequence of
the proximity of exact chiral symmetry in parameter (quark mass) space,
but it is not a true phase transition in the thermodynamic sense.

\section{Cosmological Constraints}

Considerations of phase transitions in the early Universe have been
very useful in restricting particle theory models. We now ask whether
cosmology places any constraints on the parameters characterizing the
effective potential in eq. (1).  As mentioned earlier, we will be
considering only $\epsilon_1$ and $\epsilon_2$ to be non-zero as
this is sufficient to capture the qualitative aspects of the
effective potential. It is well-known that when the effective
potential has more than one disconnected minima then domain walls
are produced in a phase transition. For the potential term in eq. (4)
this happens if $\epsilon_1 = 0$ and $\epsilon_2 \neq 0$.
Two regions of space which correspond to the two degenerate vacua
$\sigma = f_\pi$ and $\sigma = - f_\pi$, respectively, (see eq. (14)),
will be separated by a domain wall where the chiral field
smoothly interpolates between the two vacua. In the context
of the early Universe, stable domain walls are almost always disastrous
unless the phase transition happens at extremely late times.
This already suggests that the parameter $\epsilon_1$ cannot be
identically zero.

By expanding $\sigma$ and \mbox{\boldmath $\pi$}, as in eqs. (9-10),
we can determine the effective Lagrangian for $\theta$ from
eqs. (1) and (20) to be
\begin{equation}
{\cal L}_\theta = {v^2 \over 2} (\partial_\mu \theta)^2 + \epsilon_1 v
\cos\theta + {\epsilon_2 \over 2} v^2 \cos^2\theta \, .
\end{equation}
Here $v$ minimizes the effective potential at fixed $\theta$ and $T$;
that is, it traces the bottom of the valley of the potential.  Since
this valley has an almost constant radius at fixed $T$, we
can neglect its very weak dependence on $\theta$.  Hence
$v = v(T)$.

As we will show in the following, cosmology places a lower limit
on $\epsilon_1$ which is very small.
Thus, as far as cosmological considerations are concerned, we can
determine the structure of the domain wall by taking $\epsilon_1 = 0$.
With this, and defining $\theta^\prime \equiv 2 \theta$, we
get the following equation of motion for $\theta^\prime$,
\begin{equation}
\Box \theta^\prime + \epsilon_2 \sin \theta^\prime = 0 \, .
\end{equation}
This is the familiar sine-Gordan equation which is known to have
the domain-wall solutions \cite{avae}
\begin{equation}
\theta^\prime(z) \,=\, 4 \tan^{-1} \left[ \exp\left({\sqrt
\epsilon_2} z \right)\right] \, ,
\end{equation}
where the $z$ axis is normal to the wall. The thickness of
the wall is $\delta \simeq \epsilon_2^{-1/2}$.  Thus the surface
energy density $\rho_S$ of the wall is of the order
\begin{equation}
\rho_S \simeq \epsilon_2 v^2 \delta = v^2 {\sqrt \epsilon_2} \, .
\end{equation}

When $\epsilon_1$ is nonzero then, even though domain walls still
form, they are not stable any more. [It is simple to check
that domain walls always form as long as $\epsilon_1 <
f_{\pi} \epsilon_2$, which is just the condition
that there be a local minimum
at $\theta = \pi$.]  Instability of the wall arises because now one
minimum ($\theta = 0$) is energetically preferred over the other
($\theta = \pi$) so the region  corresponding to $\theta = \pi$
shrinks while the region corresponding to $\theta = 0$
expands. These unstable domain walls
then disappear in the course of time as the true minimum spreads
throughout space. Of course, in the context of the early Universe,
there is an upper bound on the life-time of such walls if they
are not to dominate the energy density of the Universe. As the
standard theory of nucleosynthesis is in very good agreement with
observations, one would also like that any such unstable domain
walls do not influence it.

A restriction on the life-time of the unstable walls implies
restriction on the parameters $\epsilon_1$ and $\epsilon_2$,
which we now consider. [Other parameters in eq. (1)  are not
constrained by such considerations from cosmology; the only other
topological objects in the model are Skyrmions, which are supposed
to be nucleons.]  Let us refer to the difference in the
energy densities at $\theta = \pi$ and $\theta = 0$
as $\Delta \rho_V$.  From eq. (23) we have $\Delta \rho_V = 2\epsilon_1 v$.
One expects that the instability of domain walls becomes
significant when the age of the Universe
$t^*$ is such that the energy excess at the scale $t^*$ becomes
comparable to the energy of the domain wall on the same
scale \cite{av}.

A good way to understand this is to realize that when
$\epsilon_1 \neq 0$ one is actually considering a situation
similar to a first order phase transition. As domain walls keep
intersecting each other and pinching off, one can consider
closed domain walls at any given time. The whole
region then looks like bubbles of one phase embedded in the
other other. The bubbles which have $\theta = \pi$ inside will always
shrink, while the bubbles which have $\theta = 0$ inside should
expand, but only
if the size of such bubbles is larger than a critical size $R_c$.
This critical size can be determined by considering the total
energy $E_R$ of a bubble of radius $R$,
\begin{equation}
E_R = - {4\pi \over 3} R^3 \Delta \rho_V + 4\pi R^2 \rho_S \, .
\end{equation}
The critical size $R_c$ is
determined by requiring that $E_R$ be stationary with respect to small
variations in $R$. Bubbles larger than $R_c$, which have
$\theta = 0$ inside, will expand
while those smaller than $R_c$ will collapse. We find that
\begin{equation}
R_c = {2\rho_S \over \Delta \rho_V} \,=\,
v {{\sqrt \epsilon_2} \over \epsilon_1} \, .
\end{equation}

So far we have neglected the fact that the Universe is expanding.
If $R_c$, at a given time, comes out larger than the horizon size,
at that time, then all the bubbles with sizes smaller than the horizon
will shrink. Bubbles with sizes equal to or greater than the horizon
always stretch along with the horizon due to the expansion of the
Universe, irrespective of which phase is enclosed.
Therefore, in order that the difference between the global minimum
and the metastable minimum becomes important, $R_c$ must be smaller than the
horizon size $H$ (corresponding to the age of the Universe at time
$t^*$).  Using the above equation, this implies that
\begin{equation}
H \,\geq\, v {{\sqrt \epsilon_2} \over \epsilon_1} \, .
\end{equation}
Using the constraint given by eq. (18) we can write this as
\begin{equation}
H \,\geq\,  {v{\sqrt \epsilon_2} \over f_\pi (m_\pi^2 - \epsilon_2)}
\,=\, \frac{v\sqrt{m_{\pi}^2-\epsilon_1/f_{\pi}}}{\epsilon_1} \, .
\end{equation}

The horizon size at the time the Universe passed through the
chiral/confinement phase transition/crossover is about $10^6$ cm,
which is {\em very} large compared to QCD scale of $10^{-13}$ cm.
Unless $\epsilon_2$ is {\em extremely} close to $m_\pi^2$ the
above inequality will be trivially
satisfied. [Note that the domain walls do not really have to
disappear much before the time of nucleosynthesis.  Therefore the
real constraint is somewhat weaker than this.]
Since $v \leq f_{\pi}$ we find the lower bound to be
\begin{equation}
\epsilon_1 > 3 \times 10^{-13} \;\; {\rm MeV}^3 \, .
\end{equation}
As long as $\epsilon_1$ is larger than this
the domain walls will disappear very quickly and will not affect
the Universe in any significant way.  Because the horizon is very small
at that time, any density fluctuations generated by collapsing
domain walls will also get wiped out quickly.

Clearly, the constraint on $\epsilon_1$ given by eq. (31)
(and the corresponding constraint on
$\epsilon_2$) is extremely weak.  Since eq. (1)
describes an effective theory anyway, it is safe to say that
cosmology imposes no practical constraints on the parameters of this model.

\section{High Energy Nuclear Collisions}

We now consider chiral symmetry breaking in the context of
quark-gluon plasma formation in a heavy ion
collision and the influence of misaligned chiral condensate
on pion production.  As we mentioned in the introduction,
the possibility of coherent pion emission from extended domains is very
interesting and has been a subject of many investigations
recently \cite{bkt,bd,rw,ggp,lrg}.
One of the problems in getting a clean signature is that such domains
are expected to be very small \cite{ggp}.
However, these investigations have been restricted
to the case when only $\epsilon_1$ is non-zero. As we have discussed
earlier, there does not seem any reason to exclude other symmetry
breaking terms in the potential. The structure of
the system drastically changes when we consider $\epsilon_2$ also
non-zero, as exemplified by the presence of domain walls.
In this section we will consider what sorts of signatures one can
expect for these more general possibilities for the symmetry breaking
terms in the effective potential.

Let us first consider pion production due to different regions of misaligned
condensate.  Since previous studies have considered only $\epsilon_1$
non-zero, we first briefly comment on this case.
It has a uniformly tilted potential with
unique minimum at $\sigma > 0$. It has been suggested earlier
\cite{rw} that, in a rapid phase change, the chiral field could roll
down to different minima in different regions. In a given region, a pion
condensate could form if the chiral field points in a direction
different from the true minimum. As this pion field relaxes to
the true minimum it will lead to coherent emission of pions.

One difficulty with getting a clean signal in this
scenario is that the typical domain one
expects is very small, of order 1 fm \cite{ggp}. However, we would
like to point out that this does not exclude the possibility of
the formation of pion condensate in a large region. For example,
consider two adjacent domains where the
chiral field points in two different directions, say $\theta_1$
and $\theta_2$.  As the chiral field relaxes, one may expect that
both $\theta_1$
and $\theta_2$ will approach zero. However, this really depends on
the values of these angles. For example, assume that $\theta_1 = \pi
+ \alpha_1$ and $\theta_2 = \pi - \alpha_2$, where both $\alpha_1$
and $\alpha_2$ are small. Then in the region where the two domains
are in contact, the chiral field will have to smoothly interpolate
between the two angles, and hence somewhere in that region it will
point in the direction $\pi$. As the outer regions of the domains
relax to $\theta$ = 0, the chiral field may start to cover a larger
region of the order parameter space ${\cal M}$. [We use ${\cal M}$
to denote the manifold defined by the bottom of the valley traced
by the minimum of the effective potential.]
This means that the evolution of the chiral fields
in a collection of domains are not totally uncorrelated.
Essentially, the chiral field defines a smooth map
from the region covered by the domains into the manifold ${\cal M}$.
The image of this map is actually a smooth and connected patch
in the manifold ${\cal M}$.  Smoothness of this patch is enforced by the
condition that the chiral field must smoothly interpolate in-between
any two adjacent domains. When the chiral field relaxes, this patch
deforms as its portions slide down to the global minimum. It is quite
conceivable that this type of evolution of the chiral field
leads to a pion condensate pointing in a direction which is in some
way the average direction defined by the patch. It is thus
not clear that small individual domains imply no
pion condensate in large regions.

We will not, however, pursue this line of argument since
$\epsilon_2 = 0$ is unnecessarily restrictive.
Rather, we will show that a non-zero value of $\epsilon_2$ leads
to the existence of large domains in a very natural manner.

As discussed in sect. 2, non-zero values of $\epsilon_2$ lead to
the existence of domain walls if the inequality $\epsilon_1
< f_\pi \epsilon_2$ is satisfied. [If not then
there is no qualitative difference between this case and the
case with $\epsilon_2 = 0$.]  These walls are unstable for all
non-zero values of $\epsilon_1$.  Consider the chiral
phase change in a region of the quark-gluon plasma. As now
there are two minima, a local minimum at $\sigma = v(\pi)$
and the global minimum at $\sigma = v(0)$, the chiral field
can relax to either of these two values. One therefore expects
a domain pattern as shown in figure 5a. For pion production,
the initial size of these domains is not crucial to our model;
this clearly distinguishes our case from previous considerations,
where the size of the initial domain was crucial.
In figure 5a we have denoted different domains by the angle
to which the chiral field relaxes in that region.

Initially, when the temperature is high, the global minimum
is only very slightly preferred over the local minimum. At those
early times, small walls may collapse
but large ones may simply be stretched by the expansion of
the plasma.  This is very similar to the situation in the early Universe.
As the plasma cools, the energy density
decreases, eventually reaching a point when the walls become unstable
in the sense that the $\theta = 0$ minimum becomes favorable over
the $\theta = \pi$ minimum.
If the expansion is slow then all the walls with sizes smaller
than the critical size [as given by eq. (28)] will shrink and disappear.
In any case, one is led to a
hierarchy of sizes of collapsing walls (see figure 5b)
which will lead to emission of pions.  As the walls
collapse, the average domain size will increase. Once the typical
domain size becomes larger than about $R_c$,
the instability of walls will become significant.
After this the regions with $\theta = 0$ will expand and the
regions with $\theta = \pi$ will contract.

The simplest prediction is the formation of walls with size of
order $R_c$ carrying excess energy density. It
is important to note that these large walls will form irrespective
of the size of the initial domains. These walls
may expand, converting the false minimum into the true one, or they may
collapse if they enclose the false minimum.  One would expect
generally that a large wall (comparable to the size of the system)
may be left enclosing the true minimum as shown in figure 6.
Between this wall and the outer boundary of the plasma region
the chiral field is in the metastable phase.  As the
$\theta = 0$ phase expands both from the outside and from the inside
the two walls will meet. This will lead to a shell of the size
of the system containing high surface energy density. All
the energy contained in the two walls will be converted to pions.

{\em The most important feature of these pions is that they
ought to be emitted from a shell-like region; studies of Bose-Einstein
correlations of pions should be able to reveal such a shell structure.}
Investigations in \cite{pratt} could be useful in this context.
A second feature is the possibility that
the pions emitted from such a shell, or from the collapsing walls,
may be coherent. We now address this possibility.

Consider a closed wall which bounds a region of the false minimum
embedded in the true minimum.  This wall will collapse and all the
energy contained in the wall will be converted into pions. The
initial structure of the wall is determined by the details of
the variation of the chiral field from the $\theta = 0$ region to
the $\theta = \pi$ region. We remind ourselves that we are actually
dealing with the minimum of the effective potential which is topologically
a three-sphere $S^3$. Different portions of the wall correspond to
different trajectories which the chiral field traces from the north
pole of this $S^3$ to its south pole. Clearly there is no
reason to expect that different portions of the wall will
all correspond to the same curve on $S^3$ initially.
One can then think of the initial distribution of the chiral field
in the entire wall as a thick strip joining the north and the south poles
of $S^3$.

As the wall collapses, one will also expect the thickness of this strip
to decrease because a thick strip costs gradient energy.
Since the pion is very light, the shrinking of a strip on
$S^3$ may proceed faster than the wall shrinks, especially
for the walls which do not shrink initially, either due to
the expansion of the plasma, or due to being least unstable.
If this strip on $S^3$ shrinks
significantly before the wall completely collapses, then the entire
collapsing shell will correspond to a chiral field which, though
still interpolating between $\theta = 0$ and $\theta = \pi$, goes
through a unique plane. For example, this curve may lie entirely
in the $\pi_1 - \pi_2$ plane.  This {\em might} lead to a pion condensate,
culminating in all of the wall energy being emitted in coherent pions.
[Of course, the process of shrinking of the strip on $S^3$ also
will produce pions but they will not be coherent.]  The same consideration
can be applied to the type of situation in figure 6 where coherent
pions may be emitted from a shell-like structure.

Let us make a rough estimate of the energy contained in the walls.
As an example, take $\epsilon_2 = 0.8 \, m_\pi^2$. Then with
eqs. (18) and (28) we get the size of the critical bubble to
be $R_c \simeq 4.5 \, m_\pi^{-1}$. With the surface energy density of
the wall as given in eq. (26) the net energy contained in the wall is
about 14 GeV.  This is a very large energy which can lead to
a high multiplicity of pions.  In this estimate we have only considered
the energy of the wall, neglecting the difference in energy
of the two minima.  During the collapse (or expansion) of the wall,
false minimum energy will be converted into the kinetic energy of the
wall which should be included to get the net energy emitted in pions.

We briefly discuss the possibility that our model can also
account for Centauro-like events \cite{cntr}.  A highly
energetic cosmic ray collision
may produce a tiny bubble of false vacuum such that the bubble wall
propogates initially outwards due to the initial momentum (or may be due
to the initial expansion of the partons). Eventually this
wall will collapse back. As this wall first expands and then collapses,
there may be enough time to develop a pion condensate on the wall (due
to shrinking of the strip on $S^3$ in the sense described
above), and hence lead to the emission of coherent pions.
This is especially likely as the initial size of the bubble may
be very small so the pion field configuration in its wall may be pretty
much uniform any way. Another possibility is that as the bubble of false
vacuum expands due to initial wall momentum, a true vacuum bubble of
critical size nucleates inside it. This then may lead to a large spherical
shell containing  high surface enrgy density
(similar to that in figure 6).
All this energy may then be emitted in coherent pions.

If we assume that all of the coherent pions come from a $\theta = \pi$
bubble, which eventually collapses, then the energy emitted in pions
$E_\pi$ can be related to the radius $R_F$ of this bubble using
eq. (27) as
\begin{equation}
E_\pi = 4 \pi f_\pi [f_\pi \sqrt{\epsilon_2} R_F^2 + {2 \over 3}
\epsilon_1 R_F^3] \, .
\end{equation}
$R_F$ here represents the radius of the bubble at its
largest size. The entire event will include any hadrons
produced when the bubble was nucleated as well as
the hadrons emitted at the end when the bubble completely collapses.
Presumably coherent pions will be emitted at the end of the event.
If we assume a relation between  $R_F$ and the duration of the
event $\tau$, say $R_F \sim \tau^a$, where $a$ is some parameter,
then our model predicts a very specific dependence of $E_\pi$
on $\tau$ (given by eq. (32)). If the information of $E_\pi$
and $\tau$ is experimentally available, then this equation can
be fitted with data to check our model and hopefully get the parameters
$a$, $\epsilon_1$ and $\epsilon_2$.

\section{Conclusion}

Our main idea is that the effective potential, or free energy, resulting
from chiral Lagrangian models of QCD {\em may} have a second metastable
minimum at a chiral angle of $\pi$.  We illustrated this in the linear
sigma model with two quark flavors.  Our numerical examples were
restricted to a symmetry breaking potential which had terms linear and
quadratic in sigma but, in general, there is no reason to think that
the other possible terms are ignorable.
There does not seem to be any fundamental reason why one should not
take this situation seriously.  Indeed, three flavor models with
linear plus bilinear terms were investigated briefly in the 1970's.
The motivation then was to investigate the pattern of symmetry breaking,
the analytic behavior of observables as the symmetry breaking parameters
were sent to zero, and to obtain improved phenomenology.  It should be
kept in mind that the parameters in a two flavor model can be renormalized
by the presence of a heavier, third flavor.  Generalizations of
our study to three flavors should be done.

Cosmology places a constraint on the height of a possible second minimum
relative to the true minimum.  This constraint arises from the requirement
that the energy in domain walls not upset standard calculations of
nucleosynthesis.  The constraint is so weak that it has no practical
consequences for high energy particle or nuclear experiments.

High energy nuclear collisions seem to present a remarkable opportunity
to study the topography of the effective chiral potential at finite
temperature.  Nonzero up and down quark masses spoil the ideal chiral
symmetry and smear out the probable second order phase transition.
This may be a cloud with a silver lining if a second metastable
minimum exists as it could mimic the effects of a {\em first} order
phase transition.  We have argued that formation and evolution of
domains, with their attendent domain walls, can plausibly lead to
observable consequences.  These include coherent pion emission and
Bose-Einstein interferometry of shell structures.
Detailed predictions with a specific effective
potential require numerical simulations as well as the inclusion
of vector mesons.

\section*{Acknowledgements}

This work was begun during the program {\em Strong Interactions at
Finite Temperature} at the Institute for Theoretical Physics in the
fall of 1993.  It was supported by the U.S. Department of Energy under
grant number DOE/DE-FG02-87ER40328 and by the U.S. National Science
Foundation under grant number PHY89-04035.  We thank J. D. Bjorken,
D. K. Campbell, B. Holstein, R. D. Pisarski, J. Polchinski,
J. Schechter and K. Rajagopal for discussions.

\newpage
\section*{Figure Captions}

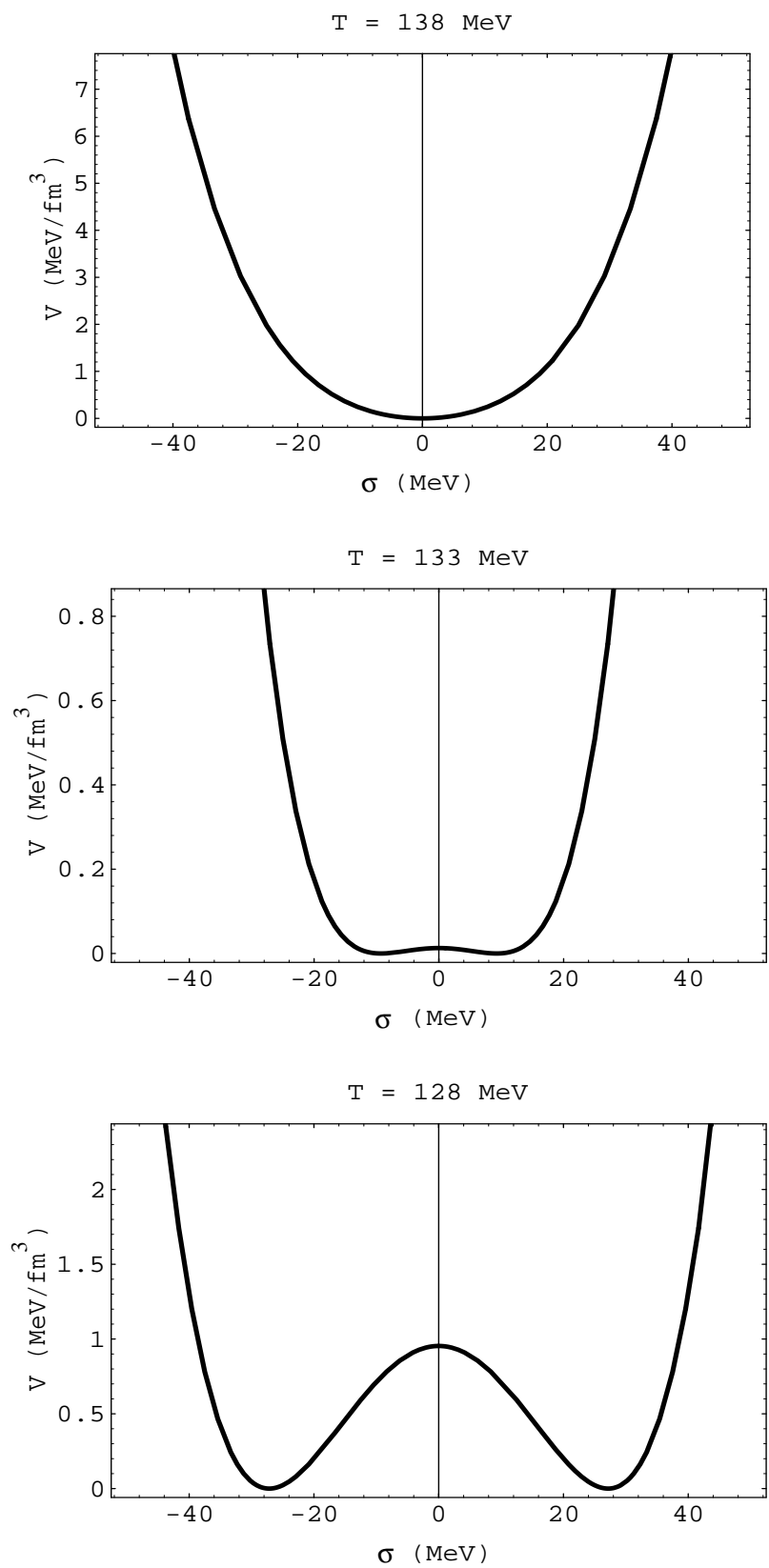
Figure 1: Temperature dependence of the effective potential for
the choice of parameters $\epsilon_1 = \epsilon_2 = 0$.
The curve shown represents a slice through the $V - \sigma$ plane.
The potential is rotationally symmetric.  There is a second order
phase transition at $T_c = \sqrt{2} f_{\pi} = 133.6$ MeV.

\vspace{.15in}

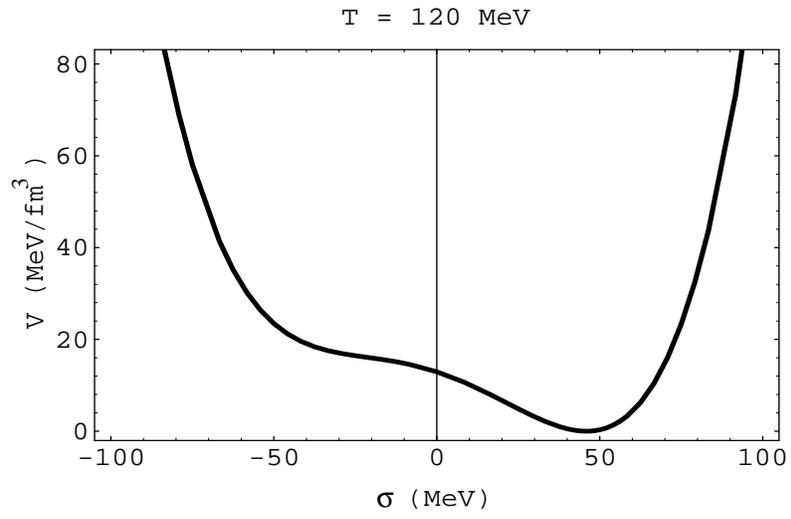
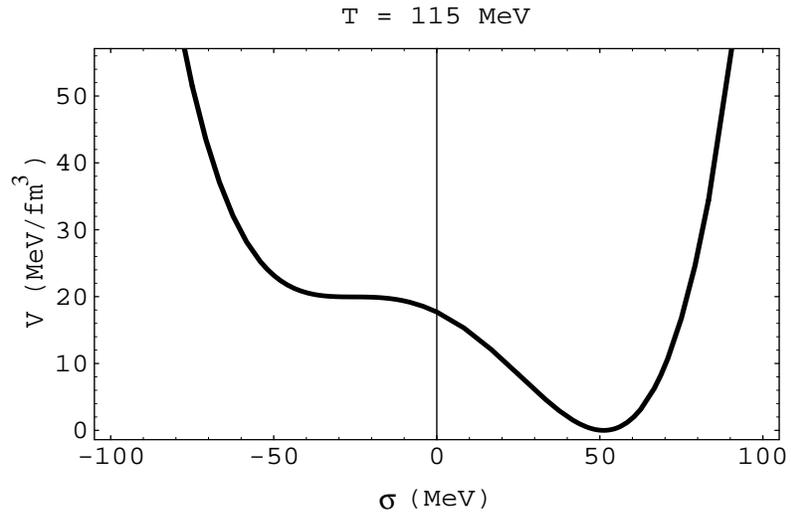
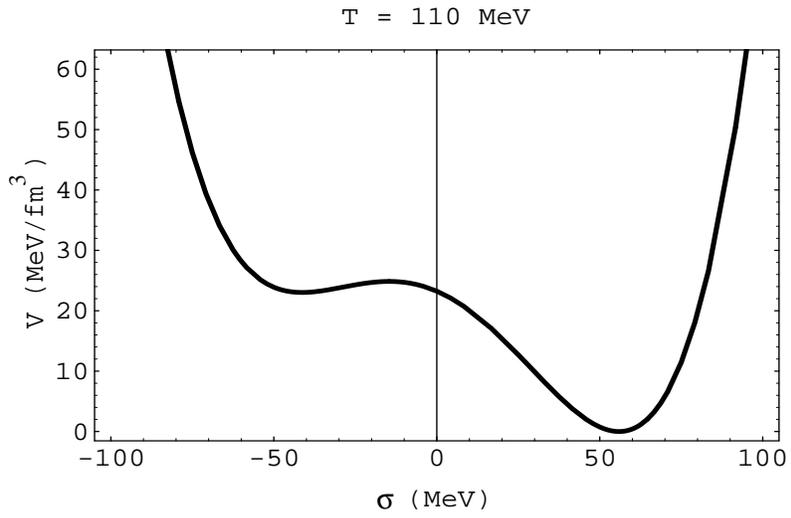
\noindent Figure 2: Temperature dependence of the effective potential for
the choice of parameters $\epsilon_1 = f_{\pi} m_{\pi}^2$ and
$\epsilon_2 = 0$.
The curve shown represents a slice through the $V - \sigma$ plane.
The bottom of the potential is tilted.  There is a saddle point
at $\theta = \pi$ for $T < 114.9$ MeV.

\vspace{.15in}

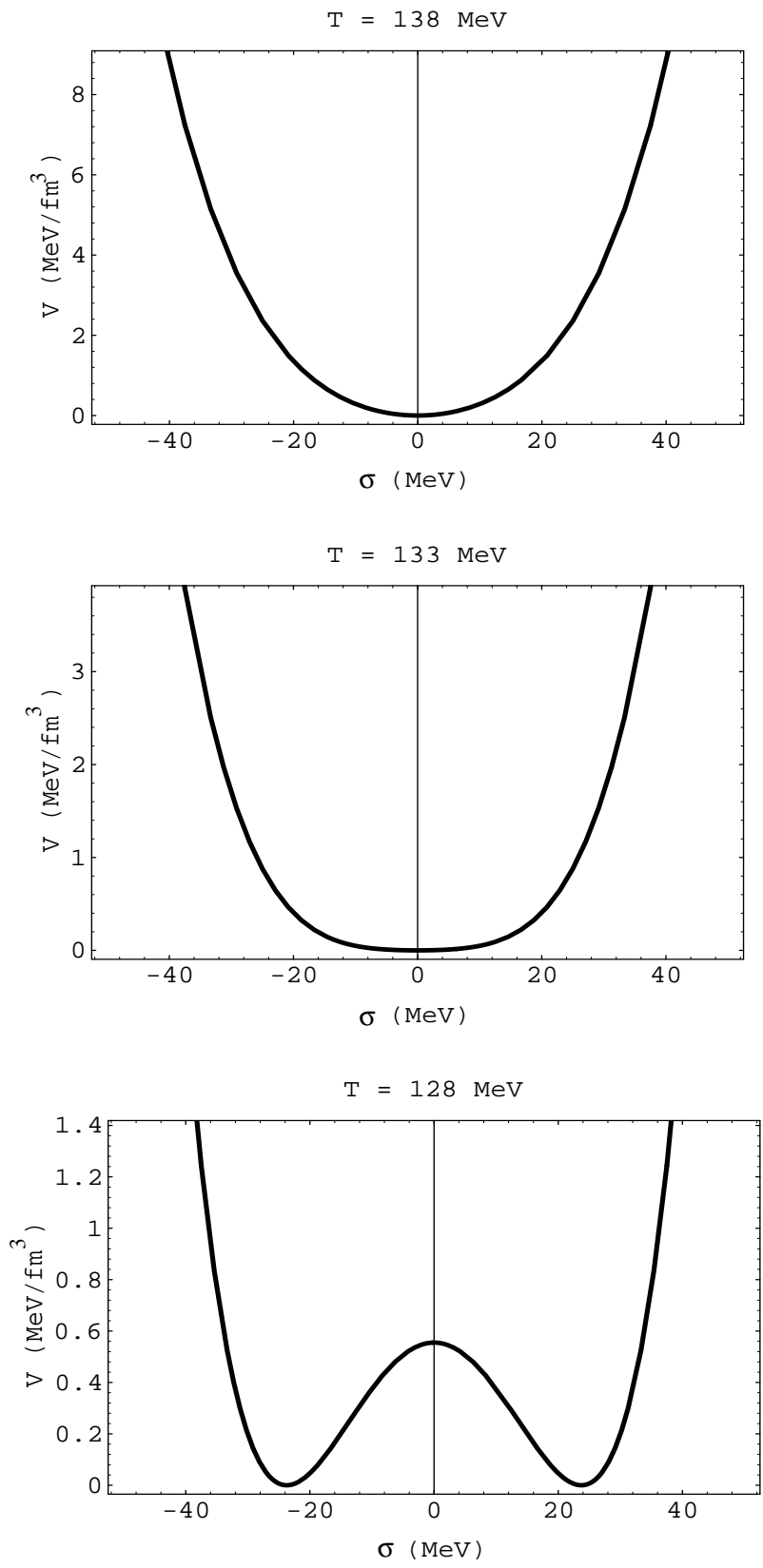
\noindent Figure 3: Temperature dependence of the effective potential for
the choice of parameters $\epsilon_1 = 0$ and
$\epsilon_2 = m_{\pi}^2$.
The curve shown represents a slice through the $V - \sigma$ plane.
There is a second order phase transition restoring the discrete
symmetry $\sigma \rightarrow -\sigma$ at
$T_c = \sqrt{2} f_{\pi} = 133.6$ MeV.

\vspace{.15in}

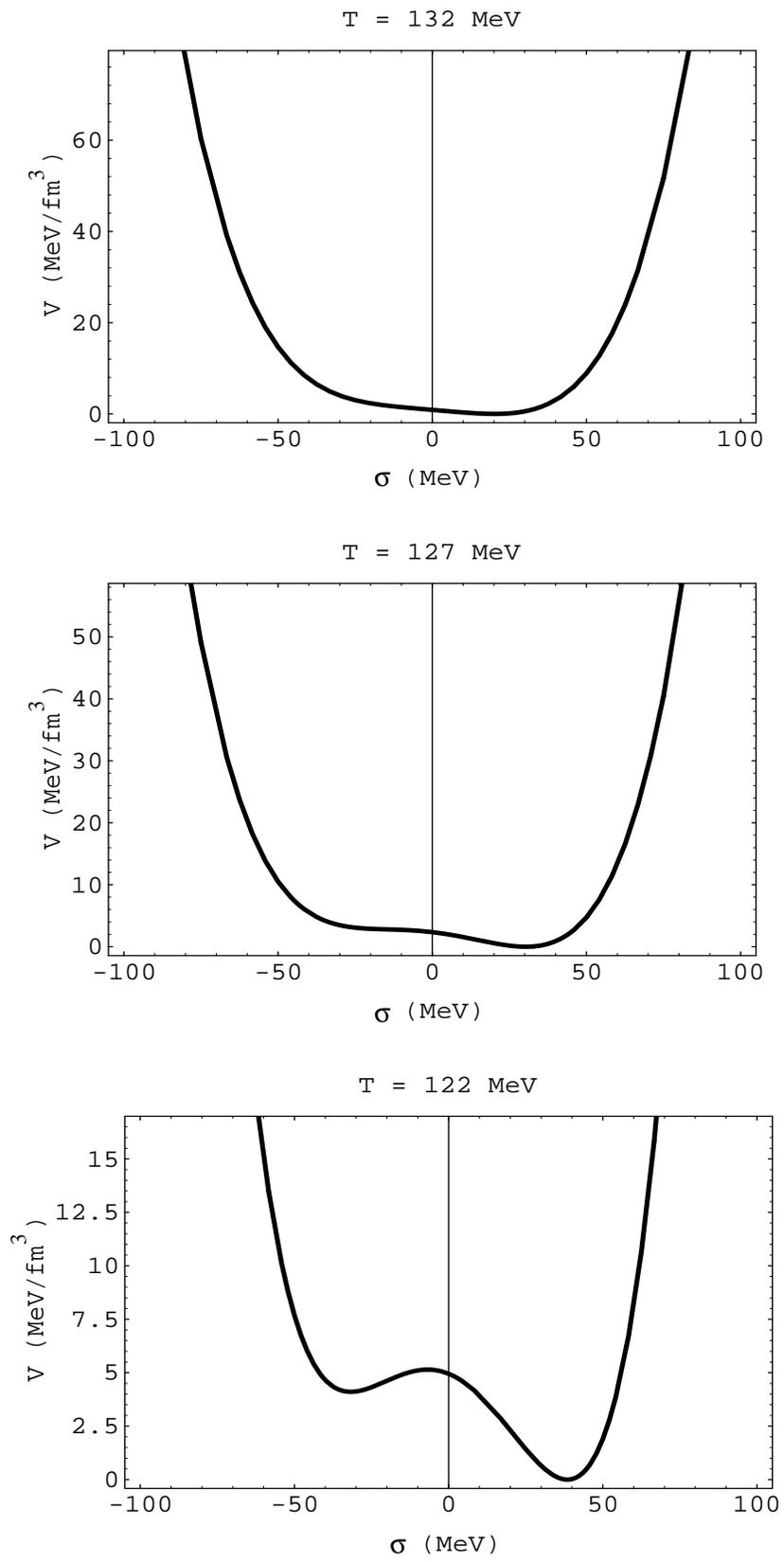
\noindent Figure 4: Temperature dependence of the effective potential for
the choice of parameters $\epsilon_1 = 0.25 \, f_{\pi} m_{\pi}^2$ and
$\epsilon_2 = 0.75 \, m_{\pi}^2$.
The curve shown represents a slice through the $V - \sigma$ plane.
The direction $\theta = \pi$ has a local minimum when
$T < T_1 = 123.2$ MeV, a saddle point when $T_1 < T < T_2 = 127.0$ MeV,
and no critical point at all when $T > T_2$.

\vspace{.15in}

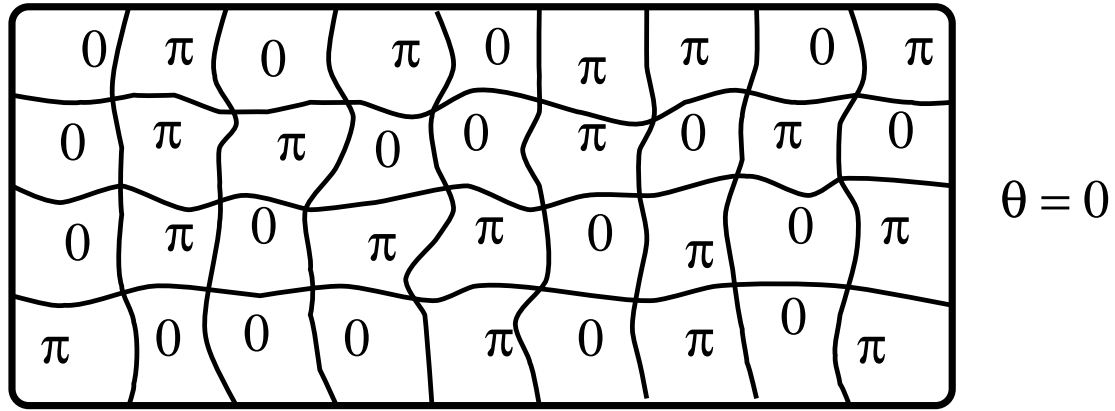
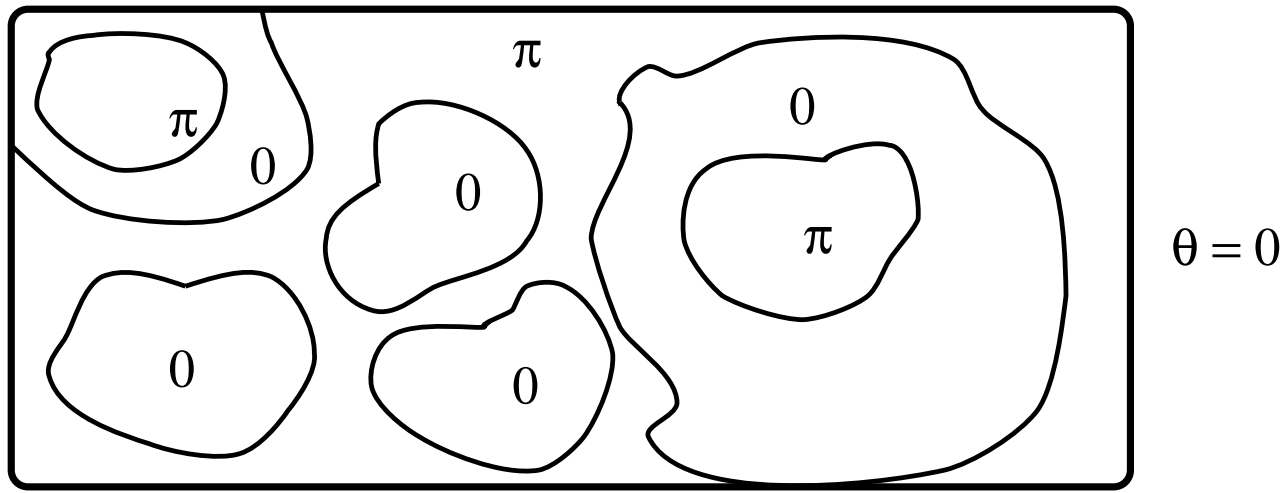
\noindent Figure 5: (a) Formation of domains after the phase transition.
Domains denoted by 0 and $\pi$ here represent regions where the chiral
field has  settled to the true and the false minimum, respectively.
Solid lines separating different domains show the initial structure
of domain walls. Outermost solid line denotes the boundary of the system.
(b) As domain walls join and collapse a hierarchy of domain sizes is generated.
The solid lines again represent domain walls separating different minima.

\vspace{.15in}

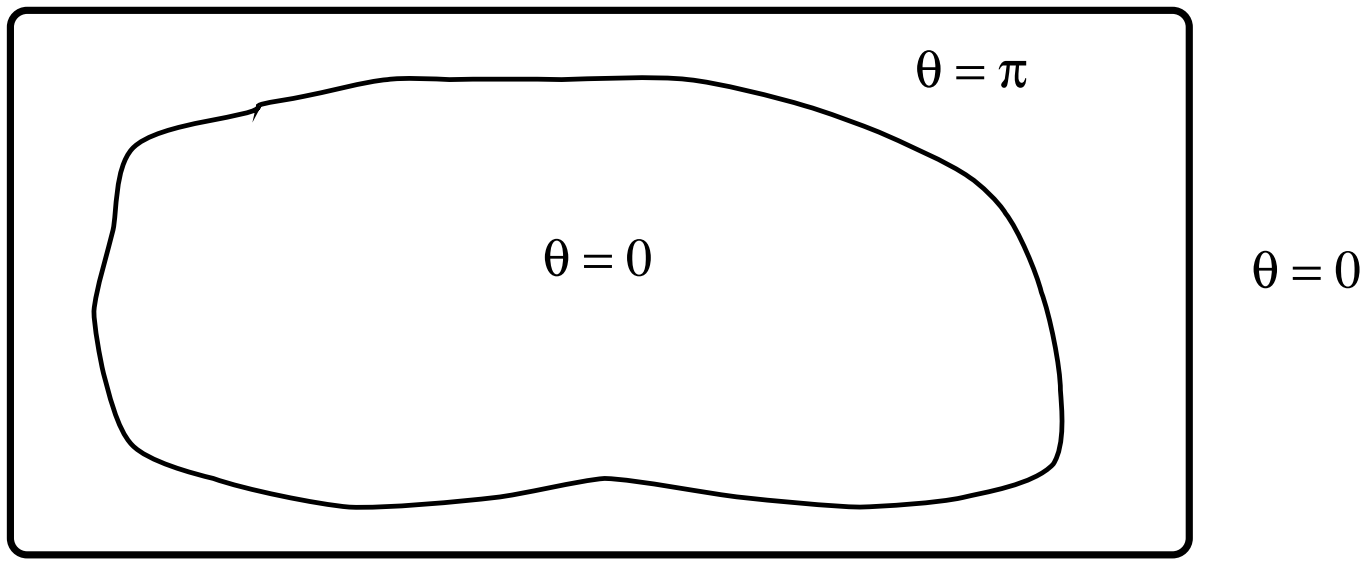
\noindent Figure 6: Eventually one may be left with a large shell-like
domain of metastable matter. As this domain shrinks, one will be left with
a large shell (of the size of the system) containing large surface density.

\end{document}